\documentclass[1, 12pt]{elsarticle}
\usepackage[english]{babel}
\usepackage[toc,page]{appendix}
\usepackage{exscale}
\usepackage{amssymb}        
\usepackage{amsthm,amsmath}        
\usepackage{natbib}         
\usepackage{mathtools}
\usepackage{graphicx}       
\usepackage{subfigure}  
\usepackage{tabularx, booktabs, multirow}    
\usepackage{color}
\usepackage{gensymb} 
\usepackage{float} 
\journal{Acta Materialia. {\rm Accepted for publication}}
\title{Influence of the stress state on the cross-slip free energy barrier in Al: an atomistic investigation} 

\author{G. Esteban-Manzanares$^{1, 2}$}
\author{R. Santos-G{\"u}emes$^{1, 2}$}
\author{I. Papadimitriou$^{1}$}
\author{E. Mart{\'\i}nez$^{3}$}
\author{J. LLorca$^{1, 2, }$\corref{cor1}}
\address{$^1$ IMDEA Materials Institute, C/ Eric Kandel 2, 28906, Getafe, Madrid, Spain. \\  $^2$ Department of Materials Science, Polytechnic University of Madrid/Universidad Polit\'ecnica de Madrid, E. T. S. de Ingenieros de Caminos. 28040 - Madrid, Spain. \\  $^3$ Theoretical Division, T-1, Los Alamos National Laboratory, Los Alamos 87545 NM, USA.}

\cortext[cor1]{Corresponding author}

\begin{document} 

\begin{abstract}
The influence of the stress state on the cross-slip rate in Al was analyzed by means of molecular dynamics simulations and transition state theory. The activation energy barrier in the absence of thermal energy was determined through the nudged elastic band method while the cross-slip rates were determined using molecular dynamics simulations for different magnitudes of the Schmid stress on the cross-slip plane, and of the Escaig stresses on the cross-slip and glide planes. The enthalpy barrier and the effective attempt frequency were determined from the average rates of cross-slip obtained from the molecular dynamics simulations. 
It was found that the different stress states influence the cross-slip rate assuming harmonic transition state theory. Moreover, the theoretical contributions to the enthalpy barrier  (configurational and due to the interaction of the applied stress with the local stress field created by the defect) were identified from the atomistic simulations  while the entropic contribution to the activation energy could be estimated by the Meyer-Neldel rule. Based on these results, an analytical expression of the activation enthalpy for cross-slip in Al as a function of the different combinations of Schmid and Escaig stress states was developed and validated. This expression can be easily used in dislocation dynamics simulations to evaluate the probability of cross-slip of screw dislocation segments. 
\end{abstract}


\maketitle

\section{Introduction} \label{Intro}

Plastic deformation in crystalline materials is mainly caused by dislocation slip along crystallographic planes. Understanding the mechanisms of plastic deformation requires a detailed modeling of the different processes that determine the dislocation motion, namely dislocation/dislocation interactions, jog formation, dislocation nucleation and annihilation, interactions of dislocations with grain boundaries and precipitates, etc. \citep{hirth1982theory, Hull2001,Nabarro2002}. In many cases, cross-slip of dislocations between different slip planes is a critical process to overcome obstacles, to create new dislocation sources or annihilate dislocations and to form dislocation patterns. In the case of FCC metals, cross-slip occurs when a screw dislocation segment with Burgers vector $b$, moving along a $\{111\}$ slip plane, changes to another slip plane which also contains the same Burgers vector. 

Dislocation cross-slip has been extensively studied and different physical processes have been proposed \citep{Puschl2002,Caillard2003}. Among them, the Friedel-Escaig mechanism \citep{Escaig1968} seems to be in very good agreement with both experiments \citep{BE79, BEM88} and atomistic simulations \citep{Kang2014}. In this process, cross-slip takes place by the creation of two Stroh constrictions along the dislocation line on the glide plane. These constrictions partially re-assemble the leading and trailing partial dislocations and form a perfect $\langle110\rangle\{111\}$ dislocation segment between them. The Burgers vector ($b$) of the dislocation segment  is contained in both the glide and cross-slip planes, and as a consequence, the dislocation is allowed to dissociate along the cross-slip plane between the Stroh nodes. 

This process is thermally-activated \citep{Caillard2003} and, hence, it depends on the magnitude of the energy barrier associated with the event. This energy barrier in FCC crystals varies with the Escaig and Schmid stresses \citep{Duesbery1992, Martinez2008, Kang2014, Malka2018, Malka2019}. The Escaig stress acts on the edge component of the Shockley partial dislocations, and controls the distance between them. The Schmid stress acts on the screw component of the dislocation and favours dislocation slip on either the glide  or the cross-slip plane.

The determination of the cross-slip rate is crucial to simulate the evolution of the dislocation network during plastic deformation and is a key ingredient of realistic discrete dislocation dynamics simulations \citep{Kubin1992,Weygand2002,Martinez2008,Hussein2015}. Thus, different approaches have been proposed in the literature to determine the energy barrier for cross-slip. Linear-elastic continuum models determine the cross-slip barrier as the extra energy that has to be supplied in order to create a perfect screw dislocation between constrictions along Shockley partials  \citep{Puschl2002,Duesbery1992,Schoeck1955, Duesbery1992II}. However, they do not deal explicitly with the forces and displacements of partial dislocations. Moreover, these models do not account for the re-dissociation contribution to the energetic barrier \citep{Puschl2002}. 

Line tension models are based on elastic interaction between Shockley partials and the energy barrier for cross-slip is obtained from the balance between repulsive interaction between partial dislocations on the glide plane and the energy to re-dissociate the partials in the cross-slip plane. Both contributions depend on applied stresses on slip and cross-slip planes. \color{black} The contribution of the stress tensor on cross-slip has been analyzed in several investigations. Kang \textit{et al.} \citep{Kang2014} analyzed the cross-slip process using approaches based on both the isotropic line tension model and atomistic string method. They reported that both methodologies provided equivalent results from the qualitative viewpoint but the influence of the applied stress was more important in the atomistic models, as compared with line tension models. Based on this work, Liu \textit{et al.} \citep{Liu2019} developed a variational line tension model for cross slip, which was in better quantitative agreement with respect to the results based on nudged elastic band model. Recently, Malka-Markovitz and Morderhai \citep{Malka2018,Malka2019} proposed a line tension model to describe the cross-slip energy barrier as a function of the stress state, which was also in in agreement with previous atomistic results \citep{Oren2017}.\color{black}

Atomistic approaches use path sampling methods (such as the nudged elastic band method and the string method) to determine the minimum energy path between two stable (or metastable) configurations to determine the energy barrier. The initial and final stable configurations are normally obtained by means of molecular statics \citep{Rasmussen1997,Vegge2001}. These atomistic strategies have been used to determine the influence of different chemical species, such as hydrogen \citep{Wen2007} or solid solution Al atoms in Ni \citep{Du2014}, on the energy barrier for cross-slip. More recently, an atomistically based model has been proposed to compute the energy barrier for cross-slip as a function of the type and volume fraction of solute atoms  \citep{Nohring2017,Nohring2018}. Additionally, Chen {\it et al.} \cite{Chen2019} studied the effect of vacancy clusters on the cross-slip activation energy in pure Ni, employing the free-end nudged elastic band methodology. They reported an overcoming mechanism map (cross-slip  \textit{vs.} shearing) as a function of the stress and obstacle spacing.  A recent investigation examined explicitly the effect of the applied stress tensor on the cross-slip energetic barrier \citep{Kuykendall2015}. Atomistic path sampling methods along with the replica trimming algorithm were employed to investigate the influence stress tensor on the cross-slip process. In addition, molecular dynamics simulations have been used to assess the time necessary for a dislocation segment to cross-slip as function of the temperature and of the applied stress and they were successfully used to determine the cross-slip activation free energy in Cu \citep{Vegge2000} and, in combination with path sampling, to determine the influence of stress on the cross-slip rate without previous assumptions \citep{Oren2017}.  Finally, Xu \textit{et al.} \citep{Xu2017} compared molecular dynamics simulations with the multiscale concurrent atomistic-continuum (CAC) methodology, which provided comparable results with lower computational cost. 

\color{black} Nevertheless, there are not atomistic models of cross-slip that take rigorously into account the effect of coupling of the different stresses on the cross-slip rate and this was the main objective of this investigation \color{black}. To this end,  atomistic simulations, in combination with harmonic transition state theory, are used to develop simple analytical expressions of the cross-slip rate as a function of the temperature and of the applied stress state in Al. The nudged elastic band method was used to determine the energy barrier in the absence of thermal energy, while molecular dynamics simulations are used to establish the influence of the temperature and of the Schmid and Escaig stresses on the free energy barrier. In particular the coupling effects between the Schmid and Escaig stresses in the glide and cross-slip planes are explicitly treated. The different terms in the analytical expressions were identified with the different contributions to the enthalpy barrier (configurational and elastic energy due to the applied stress and polarization due to the interaction of the applied stress with the local stress field created by the defect).  Moreover, the activation entropy was estimated according to the entropy-enthalpy compensation rule. The expressions can be readily introduced in dislocation dynamics simulation to predict the cross-slip rate of screw dislocation segments. 

\section{Theoretical background}\label{TB}

Transition state theory establishes the rate at which a system in equilibrium crosses a dividing energy surface between two metastable basins as a function of the activation free energy. Harmonic transition state theory (HTST) assumes that the enthalpy and entropy barriers are independent of temperature \citep{Vineyard1957}. Under this assumption, the cross-slip rate, $\Gamma _{HTST}$,  can be expressed as a function of the dislocation length ($L$) as  

\begin{equation}
\Gamma _{HTST}  = \nu_{eff} \:\left(\frac{L}{L_{n}}\right) \: e^{-\beta \Delta H (\sigma)} 
\label{TST_eq}
\end{equation}

\noindent  where $L_{n}$ is the nucleation length, i.e. the length between constrictions to nucleate the cross-slip process and $\sigma$ stand for the applied stress. $\Delta H$ is the activation enthalpy and $\nu_{eff}$ stands for the effective attempt frequency, which is proportional to the activation entropy, $\Delta S$, given by

\begin{equation}\label{n_eff}
\nu_{eff} = \nu e^{\Delta S(\sigma)/k_b}
\end{equation}

\noindent where $\nu$ is the fundamental attempt frequency.

The activation enthalpy represents the energy barrier between the saddle point and the local minimum  and, for a process at constant applied stress, it is given by

\begin{equation}
\Delta H = H^s(\sigma_{ij})-H^m(\sigma_{ij})
\label{Gibbs_sm}
\end{equation}

\noindent \color{black} where $\sigma_{ij}$ stand for the relevant components of the stress and the indexes $s$ and $m$ denote the saddle and local minimum of the energetic landscape, respectively. 
For a system under external applied stress influence and containing a single defect, the enthalpy $H$ can be expressed as 

\begin{equation}\label{H}
H = H_{conf}+ H_{el}^{st}+H_{el}^{int}
\end{equation}

\noindent where $H_{conf}$ is the configurational energy contribution and $H_{el}^{st}$ stands for the stored elastic energy as a result the combined effect of the applied stress and the internal stress induced by the defect, which is given by \color{black}

\begin{equation}
H_{el}^{st} = \frac{1}{2}\sigma_{ij}S_{ijkl} \sigma_{kl}
\label{inter_energy}
\end{equation}

\noindent  \color{black}where $S_{ijkl}$ is the fourth order compliance tensor, characteristic of the material. \color{black} The last term, $H_{el}^{int}$, stands for the energetic contribution due to the transformation of the defect under the influence of the external stress influence. \color{black} According to Clouet {\it et al.} \cite{Clouet2018}, this latter term can be determined by means of the elastic dipole approximation, which is shown to be equivalent to the infinitesimal Eshelby inclusion, within the framework of continuum elasticity \citep{Schober1984,Puls1986,Ackland1988}. Following this work, the interaction energy contribution at constant stress can be expressed as a function of the relaxation volume tensor ($V_{ij}$) according to \citep{Stoneham1983,Puls1985,Puchala2008}

\begin{equation}
H_{el}^{int} = -V_{ij}\sigma_{ij}=-V_{ij}^0\sigma_{ij}-\frac{1}{2}\Omega_{ijkl}\sigma_{ij}\sigma_{kl}
\label{inter_energy}
\end{equation}

\noindent where $\sigma$ denotes the total stress tensor (applied and internal stresses). $V_{ij}^0=-\partial H^{int}_{el}/\partial \sigma_{ij}$ and $\Omega_{ijkl}=-\partial^2 H^{int}_{el}/(\partial \sigma_{ij} \partial \sigma_{kl})$ are two tensors, of second and fourth rank, respectively, that determine the change of the energy landscape due to the applied stress and to the interaction between the applied stress and the internal stress associated with the defect, respectively. The polarization tensor $\Omega_{ijkl}$ can be defined as the difference of curvature in the free energy landscape basin because of the interaction of the stress fields, which stands for the difference of the compliance tensor in the vicinity of the defect \citep{Dudarev2018, Dudarev2018II}. 

If the difference in the stored elastic energy $H_{el}^{st}$  between the saddle and the local minimum is neglected (i.e. $\Delta H_{el}^{st} \approx 0$), eq. \eqref{Gibbs_sm} can be re-written as 

\begin{equation}
\Delta H = \Delta E_0 - \left(V_{ij}^{s,0}-V_{ij}^{m,0}\right) \sigma_{ij} - \frac{1}{2}\left(\Omega_{ijkl}^s-\Omega_{ijkl}^m\right) \sigma_{ij}\sigma_{kl}
\label{dG1}
\end{equation}

\noindent where $\Delta E_0$ stands for the free energy difference between the saddle point and the local minimum in the athermal limit due to the differences in the atomic arrangement between both configurations. Moreover, $\left(V_{ij}^{s,0}-V_{ij}^{m,0}\right)$ stand for the activation volume, which is equal to 
\begin{equation}
V^{act}_{ij} = -\partial\Delta H^{int}_{el}/\partial \sigma_{ij}
\end{equation}

\noindent multiplied by the applied stress provides the contribution of the  applied stress to the energy landscape \citep{kocks1975thermodynamics}. Thus, eq. \eqref{dG1} can be written as a function of the applied stress according to

\begin{equation}
\Delta H = \Delta E_ 0 - V^{act}_{ij} \sigma_{ij} - \frac{1}{2} \Omega_{ijkl}^{act} \sigma_{ij}\sigma_{kl}
\label{dG2}
\end{equation}

\noindent where the fourth rank polarization tensor $\Omega_{ijkl}^{act}$ enables to compute the interaction energy between the saddle and the local minimum. 

\section{Atomistic simulation methodology} \label{Methodology}

\begin{figure}[t]
	\centering
		\includegraphics[width=\textwidth]{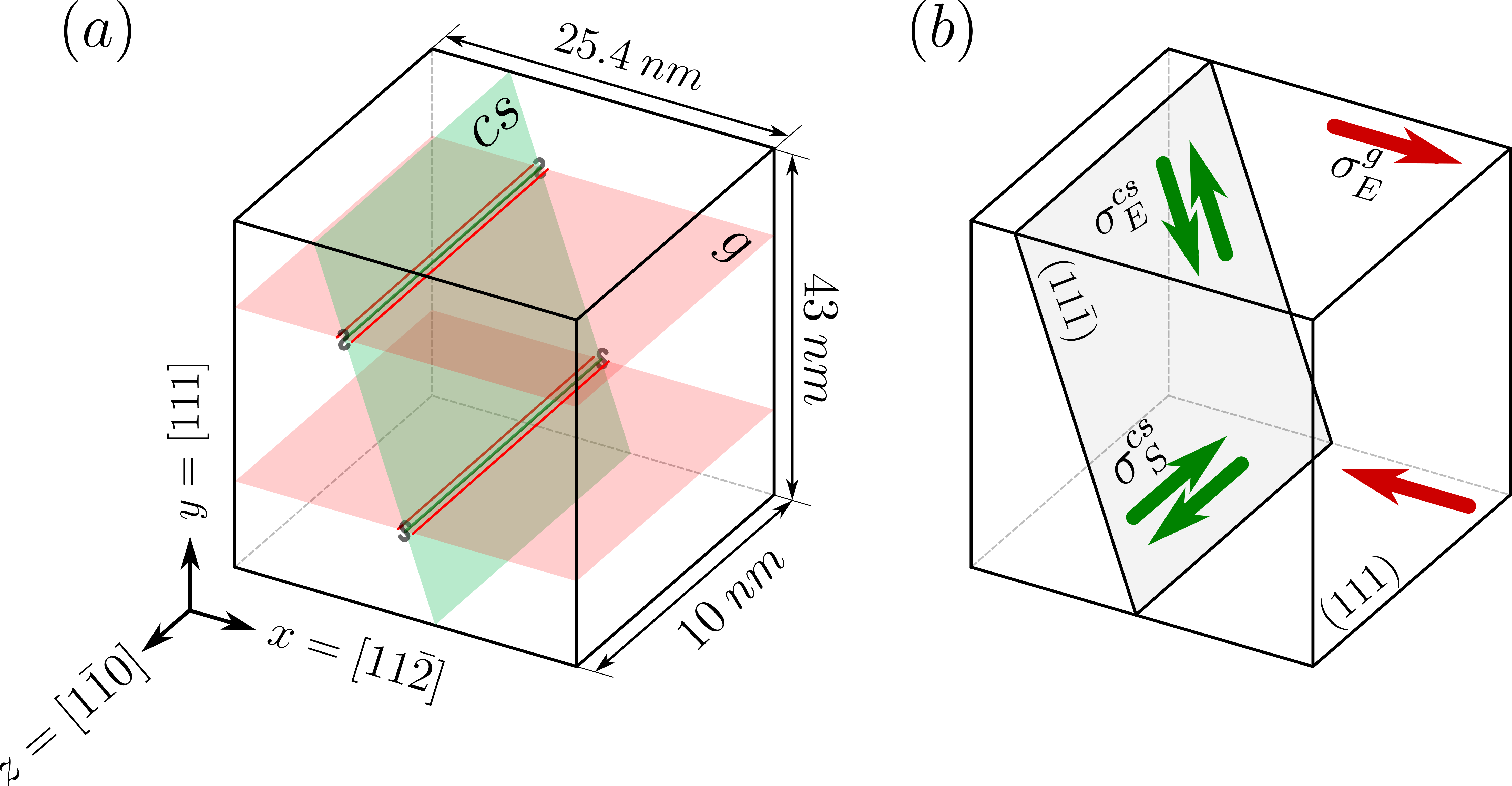}
	\caption{(a) Schematic of the atomistic domain used in all simulations. The red planes stand for the primary glide planes ($g$) of the screw dislocations while the green one is the secondary cross-slip glide plane ($cs$). Red lines depict the dissociated Shockley partial dislocations. (b) Escaig ($\sigma_E^{g}$ and $\sigma_E^{cs}$) and Schmid ($\sigma_S^{cs}$) stresses along the primary and secondary glide planes.}
	\label{Box}
\end{figure}

All atomistic simulations were carried out in a parallelepipedic box (Fig. \ref{Box}). The $x$, $y$ and $z$ axes of the box were parallel to the $[1 1 \bar{2}]$, $[1 1 1]$ and $[1 \bar{1} 0]$  crystallographic directions of the Al FCC lattice, respectively. These directions stand for the normal to the dislocation line, the normal to the glide plane and the dislocation line, in the same order. The dimensions of the domain were $25.4 \times 43 \times 10$ nm$^3$ along the $x$, $y$ and $z$ axes, respectively. Periodic boundary conditions were applied along the three axes of the box. Two screw dislocations with opposite sign and equally spaced in the $y$ axis, were inserted in two $(111)$ planes perpendicular to the $y$ axis with the dislocation line parallel to $[1 \bar{1} 0]$ ($z$ axis). The perfect screw dislocations were introduced by applying the corresponding isotropic displacement field to each atom using the open source code ATOMSK \citep{Hirel2015}. The dislocations were located along the $y$ axis, on parallel $(111)$ planes, separated 21.5 nm each other, as illustrated in Fig \ref{Box}$(a)$.  The energy of the screw dislocation dipole was minimized using the conjugate gradient algorithm and the perfect dislocations were split into two Shockley partials. The internal atomic coordinates were considered optimized once the global interatomic force was below 10$^{-8}$ eV/$\mathbb{\AA}$. The EAM interatomic potential developed by Mishin \textit{et al.} for Al was used \citep{Mishin1999,apostol2011interatomic}. 

Two different types of simulations were carried out with the optimized structure. A multi-replica static optimization strategy using the NEB method was applied to obtain the minimum energy path (MEP) and the activation energy for cross-slip in Al. Afterwards, molecular dynamics (MD) simulations were performed to obtain the enthalpy barrier and the attempt frequency as a function of the temperature and of the stress state. All the simulations were carried out using LAMMPS \citep{plimpton2007lammps}. OVITO \citep{stukowski2009visualization} was used to visualize the results. 

\subsection{Determination of the activation energy at 0K}

The standard NEB algorithm \citep{Mills1995, Henkelman2000a} along with the climbing image nudged elastic band (CI-NEB) method \citep{Henkelman2000b} were employed. The combination of both algorithms allows an accurate determination of the MEP and of the activation energy for cross-slip in Al. This static procedure is based on the energy minimization of a set of replicas of the same system located between 2 local minima along the potential energy landscape. Both the initial and final states were inputs for the analysis. In the initial state of the system, both screw dislocations lie along the glide plane ($g$). In the final state, one dislocation continues on the glide slip plane while the second dislocation is located on the cross-slip plane ($cs$). The atomic positions between both minima were obtained by linear interpolation and 32 replicas were used in the simulations. The stiffness constant of the band between replicas was set to 10 eV/$\mathbb{\AA}$. 
The fast inertial relaxation engine (FIRE) algorithm \citep{Sheppard2008} was used to optimize the path along the potential landscape with the interconnected replicas. The timestep for the dynamic calculation was 1 fs. The set of replicas was considered to be located along the minimum energy path when the overall force of each replica was below 10$^{-3}$ eV/$\mathbb{\AA}$.

\subsection{Determination of the rate at finite temperature and stress}\label{subsecMD}

MD simulations were used to analyze the cross-slip process in Al as a function of the stress state. The Schmid stress on the glide plane ($\sigma_{S}^g$) is parallel to the Burgers vector of the full dislocation and moves both partial dislocations in the same direction, while the Escaig shear stress on the glide plane ($\sigma_{E}^g$) is perpendicular to the full dislocation Burgers vector and expands or shrinks the stacking fault area between partials. Similar definitions can be given for the Schmid ($\sigma_{S}^{cs}$) and Escaig ($\sigma_{E}^{cs}$) shear stress on the cross-slip plane. Considering the  orientation of the atomistic domain illustrated in Fig. \ref{Box}, they are expressed by

\begin{equation} \label{Esc_Sch}
\begin{array}{ll}
\sigma_{S}^g=\tau_{yz} & \qquad \sigma_{S}^{cs} =\frac{2\sqrt{2}\tau_{yz}-\tau_{xz}}{3} \\
\sigma_{E}^g=\tau_{xy} & \qquad \sigma_{E}^{cs}=\frac{2\sqrt{2}\left(\sigma_{xx}-\sigma_{yy} \right)-7\tau_{xy}}{9}
\end{array}
\end{equation}

The equilibrium condition ($\sigma_{S}^{g}$ = 0) was assumed in all the MD simulations. Thus, the dislocation does not move on the primary glide plane, and the rate for dislocation cross-slip depends only $\sigma_{S}^{cs}$, $\sigma_{E}^{g}$ and $\sigma_{E}^{cs}$. Different combinations of stresses along the glide and cross-slip planes were used to analyze the process, uncoupled ($\sigma_{E}^g$, $\sigma_{S}^{cs}$ and $\sigma_{E}^{cs}$), dual coupling ($\sigma_{E}^g+\sigma_{E}^{cs}$, $\sigma_{E}^g+\sigma_{S}^{cs}$ and $\sigma_{E}^{cs}+\sigma_{S}^{cs}$) and full-coupling ($\sigma_{E}^g+\sigma_{E}^{cs}+\sigma_{S}^{cs}$). They were obtained by controlling the applied stresses $\tau_{xz}$, $\tau_{xy}$, $\sigma_{xx}$ and $\sigma_{yy}$ on the simulation box. The stress states applied to the atomistic system in order examine cross-slip are detailed in the \ref{A2}.

All the MD simulations were carried out using the NPT ensemble, in which different stress states were used to change the Escaig and Schmid stresses. The timestep employed in these simulations was 2 fs. An initial temperature/stress stabilization was carried out by increasing linearly the temperature and stress up to the chosen level during 40 ps. The time elapsed for cross-slip, $t^{MD}$, was computed from the instant at which temperature and stress attained the target levels.  $t^{MD}$  was measured until the cross-slip process was carried by one of the dislocations of the dipole. More specifically, the position of the dipole stacking faults between partial dislocations along $[11\bar{2}]$ direction was traced every 2 ps. It should be noted that the dislocation lying on the cross-slip plane moved towards the other dislocation of the dipole and annihilated immediately after the cross-slip process was completed. This effect was used as an indicator to finalize the MD simulations.  Fifteen uncorrelated simulations were carried out for each temperature/stress pair to obtain statistically significant results. The average cross-slip rate as a function of temperature and stress was calculated as

\begin{equation}
\overline{\Gamma}\left(T,\sigma\right)=\frac{2}{\left<t^{MD}\left(T,\sigma\right)\right>}
\end{equation}   

\noindent where $\left<t^{MD}\right>$ stands for the average MD simulation times for all the simulations performed for cross-slip at given temperature/stress pair and the factor 2 is due to the existence of two dislocations in the simulation box. 
For each stress state, the average rates were computed for at least two different stress values at five different temperatures in the range 400 K to 600 K in steps of 50 K. 

\section{Results and discussion} \label{Results}

\subsection{Nudged elastic band simulations results}\label{NEBsec}

\begin{figure}[!]
	\centering
		\includegraphics[width=0.9\textwidth]{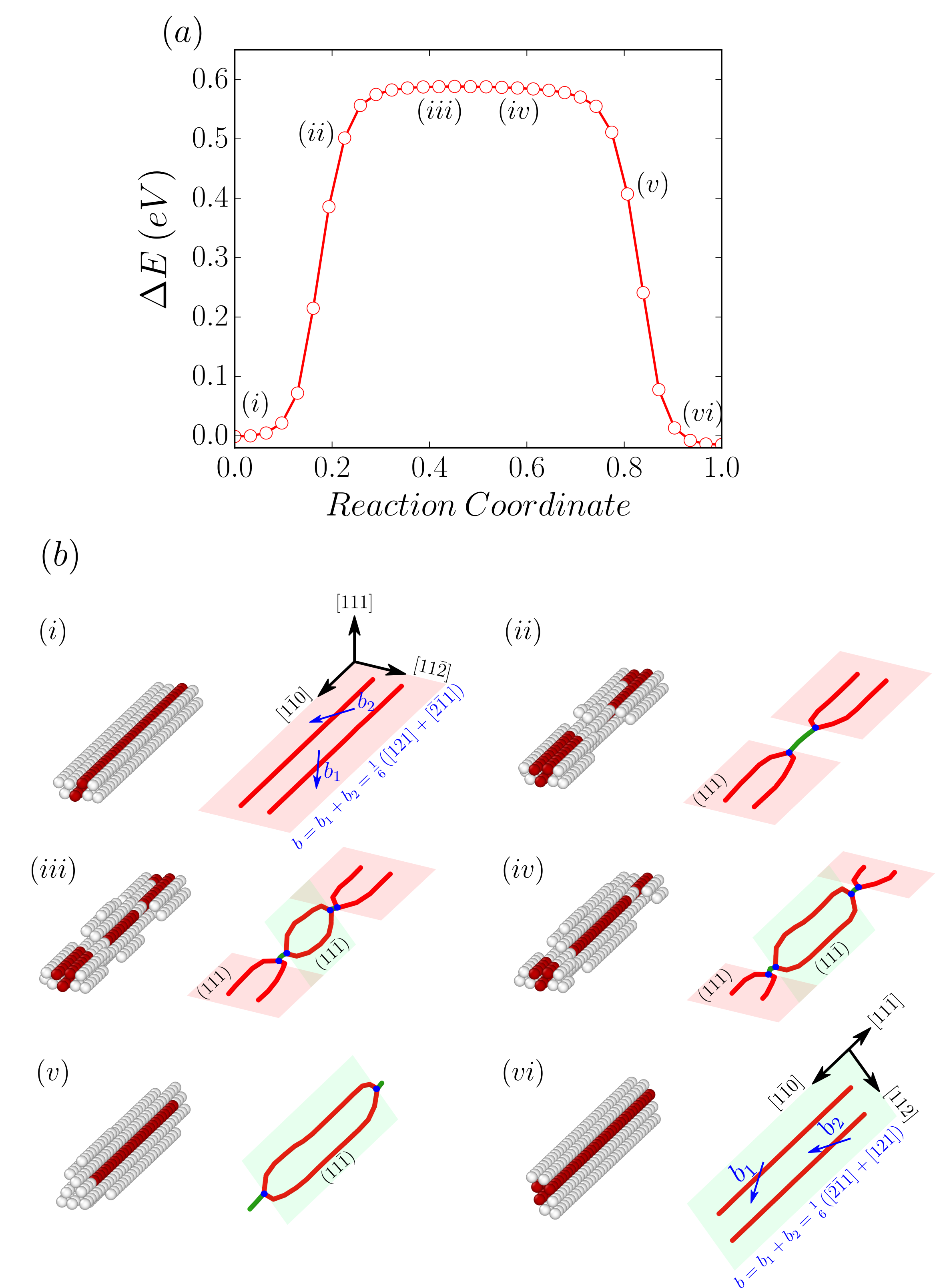}
	\caption{(a) Energy landscape along the minimum energy path for cross-slip at 0K in Al. The activation energy ($\Delta E_0$ = 0.582 eV) for the process corresponds to the saddle point of the curve. (b) Atomistic snapshots along with schematic representations of the partial dislocation lines during the cross-slip process obtained from the NEB calculations. The Friedel-Escaig mechanism is clearly observed.}
	\label{NEB_Results}
\end{figure}

The NEB calculations provided the activation energy $(\Delta E_0)$ for cross-slip at 0 K as well as the particular cross-slip mechanism that follows the MEP in absence of thermal energy. The evolution of the internal energy along the MEP is plotted in Fig. \ref{NEB_Results}$(a)$. The curve shows a smooth transition between both the initial and final minima and the activated state. The energy barrier $\Delta E_0$ given by the saddle point in the curve was 0.582 eV which corresponds to the configurational energy contribution in eq. \eqref{dG2}.

The cross-slip mechanism provided by NEB simulations is shown in Fig. \ref{NEB_Results}$(b)$. Different snapshots of the atomic positions along the dislocation line during the cross-slip process are illustrated in this figure, together with the schematic representation of the position of the partial dislocations. The process follows the Friedel-Escaig mechanism. Both Shockley partials (red lines) can be seen in their ground state on the $(111)$ plane in  $(i)$. Two constrictions, represented as blue dots, are formed in $(ii)$, near the saddle point. They are linked by a perfect [110](111) dislocation (green line). The distance between both constrictions was approximately 2.8 nm. This distance is the nucleation length, $L_{n}$, i.e. the length between constrictions to nucleate the cross-slip process. At the saddle point, depicted in $(iii)$, the perfect dislocation dissociates along the $(11\bar{1})$ plane.  Thus, the MEP for cross-slip involves the creation of two Stroh constrictions along the dislocation line tangent and the re-dissociation of the dislocation segment between the two constrictions on the cross-slip plane.

  It is worth noting that this mechanism is slightly different from Friedel observations, who considered that \color{black} the partial dislocations would immediately re-dissociate  on the cross-slip plane between the two Stroh nodes. However, our calculations show the formation of a perfect screw dislocation segment between the nodes.  This effect was reported recently by Oren \textit{et al.}\cite{Oren2017}. \color{black} Once the saddle point is reached, both constrictions expand in opposite directions, enforcing the re-dissociation of the partials on the cross-slip plane. This process is illustrated in $(iv)$ and $(v)$. Finally, both partial dislocations are completely dissociated in the $(11\bar{1})$ plane, as shown in $(vi)$. It should be noted that this mechanism was already observed in other atomistic studies of cross-slip \citep{Wen2007,Du2014,Oren2017}.

The cross-slip barrier obtained by NEB is in good agreement with the cross-slip line tension models \citep{Escaig1968}. These models claim that the activation barrier for cross-slip is defined by the contribution of three different energetic sources, i.e. the elastic repulsion energy between two partials, the change of line energy and the variation of the stacking fault due to the effect of Schmid and Escaig stresses along either glide and cross-slip planes \citep{Kang2014}. 

\subsection{Molecular dynamics results}\label{MDsec}

\subsubsection{Uncoupled stresses}\label{UncStrs}

\begin{figure}[!]
	\centering
		\includegraphics[width=\textwidth]{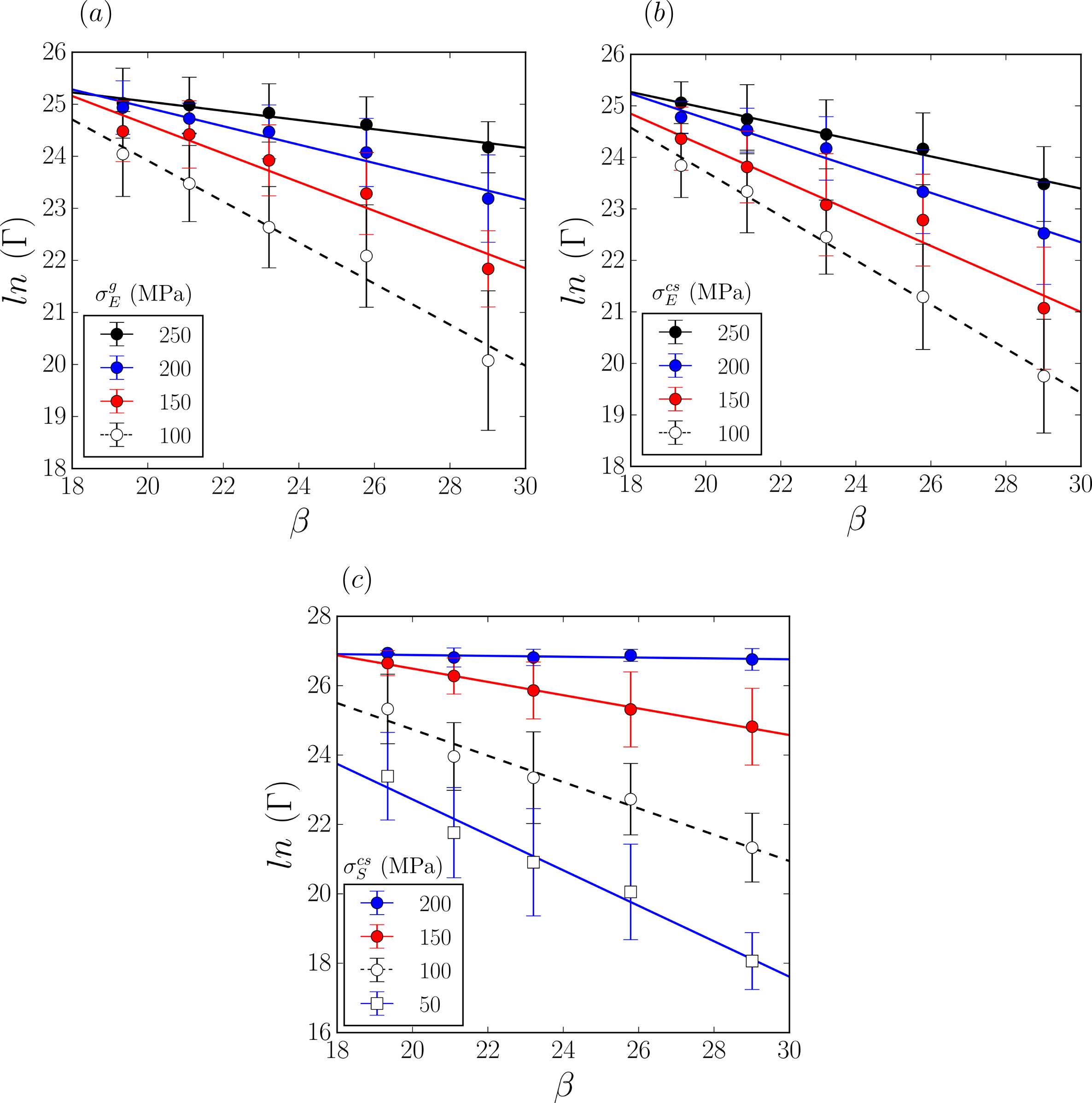}
	\caption{Average rates of the cross-slip process (expressed in $s^{-1}$) as a function of $\beta$ (= 1/$k_bT$) obtained from the MD simulations.  The bars denote the standard deviation from the average. $(a)$ Influence of the Escaig stress on the glide plane, $\sigma_E^{g}$. $(b)$ Influence of the Escaig stress on the cross-slip plane, $\sigma_E^{cs}$  $(c)$ Influence of the Schmid stress on the cross-slip plane, $\sigma_S^{cs}$.}
	\label{Single_rate}
\end{figure}

The average cross-slip rates are plotted in Fig. \ref{Single_rate} as a function of $\beta$ (= 1/$k_bT$) for different values of the Escaig stress on the glide and cross-slip planes ($\sigma_E^{g}$ in  Fig. \ref{Single_rate}$(a)$ and $\sigma_E^{cs}$ in Fig. \ref{Single_rate}$(b)$, respectively) and of the Schmid stress on the cross-slip plane, $\sigma_S^{cs}$ (Fig. \ref{Single_rate}$(c)$). It should be noted that only the Friedel-Escaig cross-slip mechanism was observed for all temperatures and stress states.

As expected, the energy barrier decreased as the stress increased and also the probability to overcome the barrier increased with the temperature. Moreover, the logarithm of the rates showed a linear dependence with $\beta$ in all cases, indicating that the energetic barrier does not depend on the temperature, and, thus, HTST holds. Moreover, it was observed that the pre-exponential factor decreased as the applied stress increased, indicating that the entropic barrier was also a function of the stress. This behavior can be rationalized through the so-called Meyer-Neldel (MN) rule \citep{Meyer1937}, which establishes that the reduction of the activation energy is compensated by a reduction of pre-exponential factor for correlated events (i.e. dislocation cross-slip at different stress) and, consequently, of the entropic barrier.

The MN rule assumes that the activation entropy is proportional to the activation enthalpy according to
\begin{equation}\label{MNR}
\Delta S(\sigma)=\frac{\Delta H (\sigma)}{T_m}
\end{equation}

\noindent where $T_m$ is the melting temperature. Different studies have demonstrated the applicability of the MN compensation rule in solid mechanics \citep{sobie2017thermal, NBW11, SR11, saroukhani2016harnessing, SW17}. 

The activation enthalpy and entropy barriers for each applied stress was obtained according to eqs. \eqref{TST_eq} and \eqref{n_eff}, respectively. The length of the screw dislocation segment in all the MD simulations was $L$ = 10 nm and $L_n$ was considered equal to the nucleation length observed in the NEB simulations.

\begin{figure}[!]
	\centering
		\includegraphics[width=0.95\textwidth]{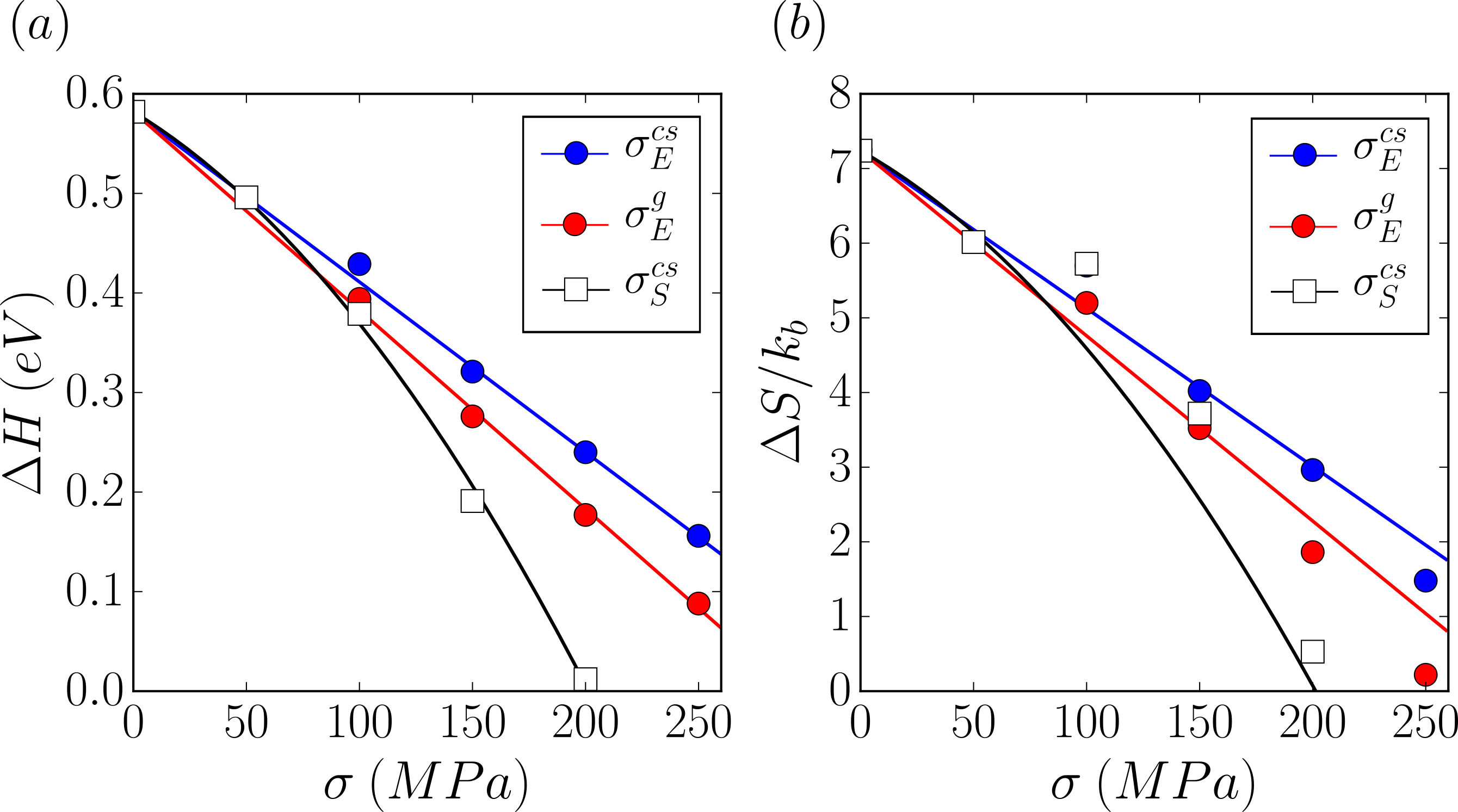}
	\caption{$(a)$ Activation enthalpy and $(b)$ activation entropy as a function of the applied Schmid and Escaig stresses on the glide or cross-slip planes. The results of the MD simulations are given by the  symbols. The lines represent the predictions of eq. \eqref{Uncoup_model} and eq. \eqref{MNR}.}
	\label{Single_G}
\end{figure}

The activation enthalpy, $\Delta H$, obtained from MD simulations is plotted in Fig. \ref{Single_G}$(a)$ as a function of the applied stress (either the glide stress on the cross-slip plane or the Escaig stress in the glide or cross-slip planes). The solid lines stand for the best fit of the MD results according to eq. \eqref{dG2}  It should be noted that a single value of the activation volume and of the interaction factor were assumed for each stress. So, the activation energy for cross-slip can be expressed as a function of the projection of the stress along the different planes according to

\begin{equation}\label{dG_expansion}
\Delta H (\sigma) = \Delta E_0 - V^{act} \sigma  - \frac{1}{2} \Omega^{act} \sigma^2 
\end{equation}

\noindent where $\sigma$ represents the corresponding stress ($\sigma_E^{g}$,$\sigma_E^{cs}$ or $\sigma_S^{cs}$).

The results of Fig. \ref{Single_G} reveal the different influence of the Escaig and Schmid stresses on the glide and cross-slip plane on the rate of the event. It can be observed that $\Delta H$ decreases linearly with the Escaig stress in both planes while the dependence of $\Delta H$ with the Schmid stress on the cross-slip plane is quadratic. Thus, the contribution of the quadratic interaction energy in the case of the Escaig stresses seems to be negligible and the enthalpy barrier can be expressed as.

\begin{equation}\label{Uncoup_model}
\begin{array}{l}
\Delta H(\sigma_E^g)= \Delta E_0 - V_E^g\sigma_E^g\\
\Delta H(\sigma_E^{cs})= \Delta E_0 - V_E^{cs} \sigma_E^{cs}\\
\end{array}
\end{equation}  

\noindent according to the MD results in Fig. \ref{Single_G}, where $V_E^g $ and $V_E^{cs} $ stand for the activation volumes for the Escaig stress on the glide and cross-slip planes, respectively. 

In the case of the Schmid stress on the cross-slip plane, this first order approximation does not suffice to explain the effect of the external stress on the energy barrier and the second order polarization term has to be included to determine the activation enthalpy according to

\begin{equation}\label{Uncoup_modelC}
\Delta H(\sigma_S^{cs})= \Delta E_0 -  V_S^{cs,0}\: \sigma_S^{cs} - \frac{1}{2}\Omega_{S}^{cs}(\sigma_{S}^{cs})^2
\end{equation} 

\noindent where $V_S^{cs}$ stands for the activation volume at zero stress and $\Omega_{S}^{cs}$ expresses the effect of $\sigma_S^{cs}$ on the activation volume due to the polarization. 

The three activation volumes and $\Omega_{S}^{cs}$ were obtained by fitting the results of the MD simulations in Fig. \ref{Single_G} to eqs. \eqref{Uncoup_model} and \ref{Uncoup_modelC} and they are depicted in the Table \ref{UncoupFit}.  It is worth noting that an important reduction of the barrier is evident in the case of the Schmid stress on the cross-slip plane with respect with the two Escaig stresses for large stresses ($\ge$150 MPa). However, the three stresses have similar influence on the energy barrier when they are below 100 MPa.

\begin{table}[h]
 \begin{center}
    \caption{Activation volumes and $\Omega_{S}^{cs}$ for uncoupled Escaig and Schmid stresses} 
    \label{UncoupFit}
    \begin{tabular}{l|c|c|c} \hline
    \hline
       & $\sigma_E^{g}$ & $\sigma_E^{cs}$  & $\sigma_S^{cs}$ \\
      \hline
        $V$ ($b^3$) & 13.6 & 11.6 & 9.4 \\
		$\Omega$ ($b^3$/MPa) & -- & -- & 0.102\\
      \hline
      \hline
    \end{tabular}
  \end{center}
\end{table}

 The effective attempt frequency $\nu_{eff}$, obtained from MD simulations, was in the range from 5$\times$10$^{11}$ to 1$\times$10$^{13}$ s$^{-1}$, which is in agreement with previous studies \citep{Vegge2000, Oren2017}. The activation entropy, $\Delta S$, was estimated from $\nu_{eff}$ and eqs. \eqref{TST_eq} and \eqref{n_eff} asumming a fundamental attempt frequency $\nu$ = 10$^{11}$ \citep{saroukhani2016harnessing, sobie2017modal}

The activation entropies, $\Delta S$, are plotted in Fig. \ref{Single_G}$(b)$ as a function of the applied stress. The symbols stand for the results obtained from the MD results while the lines define the predictions according to eq. \eqref{MNR} with $T_m$=933 K, which is the melting temperature of the Al \citep{meyrick1973phase}. It was found that $\Delta S$ decreased as applied stress increased and that this dependency was in agreement with the enthalpy-entropy MN compensation rule.

\subsubsection{Coupled stresses}

\begin{figure}[!]
	\centering
		\includegraphics[width=\textwidth]{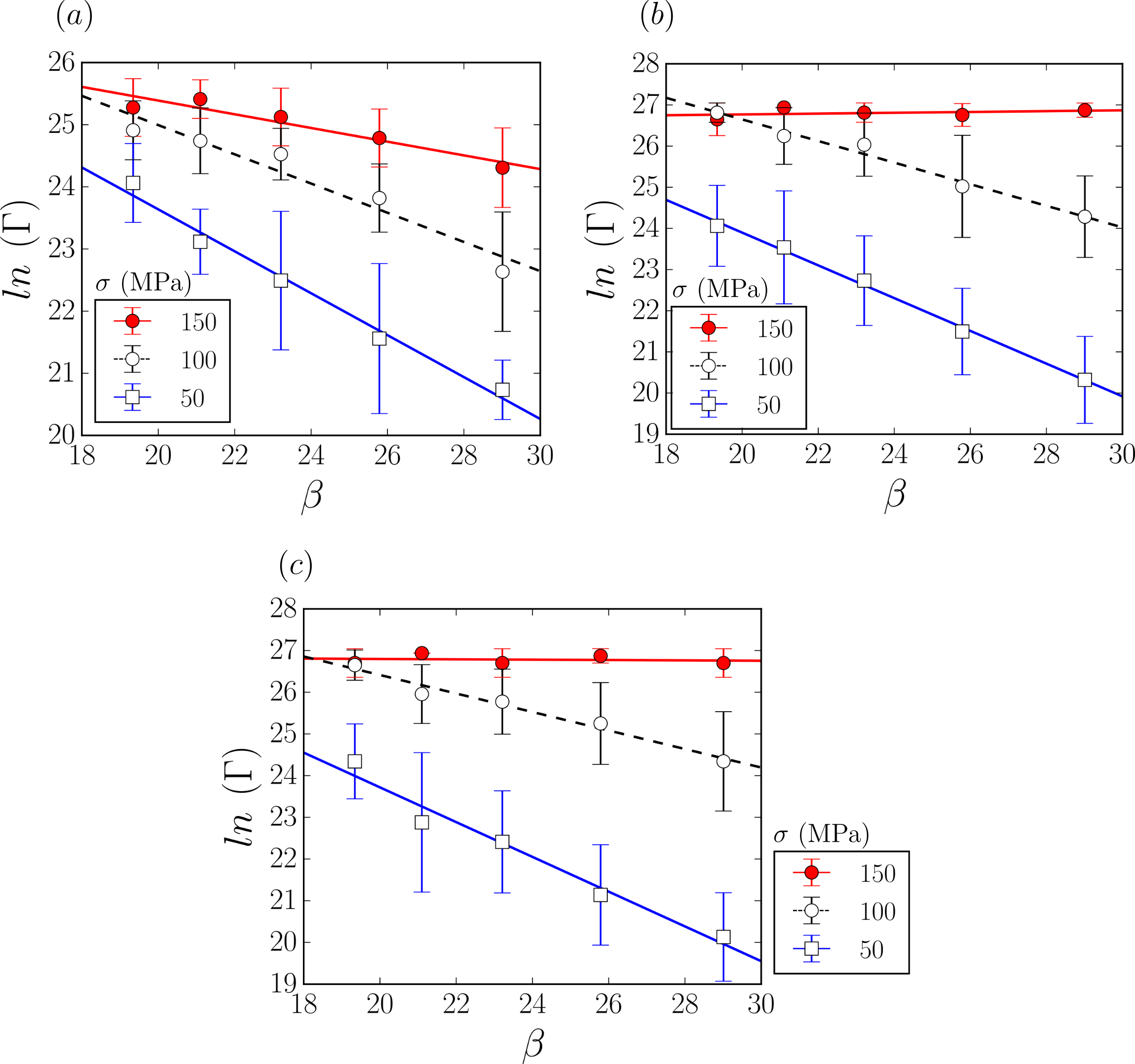}
	\caption{Average rates of the cross-slip process (expressed in s$^{-1}$) for coupled stresses as a function of $\beta$ obtained from MD simulations.  The bars stand for the standard deviation from the average rate. $(a)$ Influence of the Escaig stresses on the glide and cross-slip planes ($\sigma$ = $\sigma_E^{g}=\sigma_E^{cs}$). $(b)$ Influence of the Escaig stress on the glide plane and of the Schmid stress on the cross-slip plane ($\sigma$ = $\sigma_E^{g}=\sigma_S^{cs}$).  $(c)$ Influence of the Escaig and Schmid stresses on the cross-slip plane ($\sigma$ = $\sigma_E^{cs}=\sigma_S^{cs}$). The lines represent the fit obtained with eq. \eqref{Coup_model}.}
	\label{Coupled_rate}
\end{figure}

\begin{figure}[!]
	\centering
		\includegraphics[width=\textwidth]{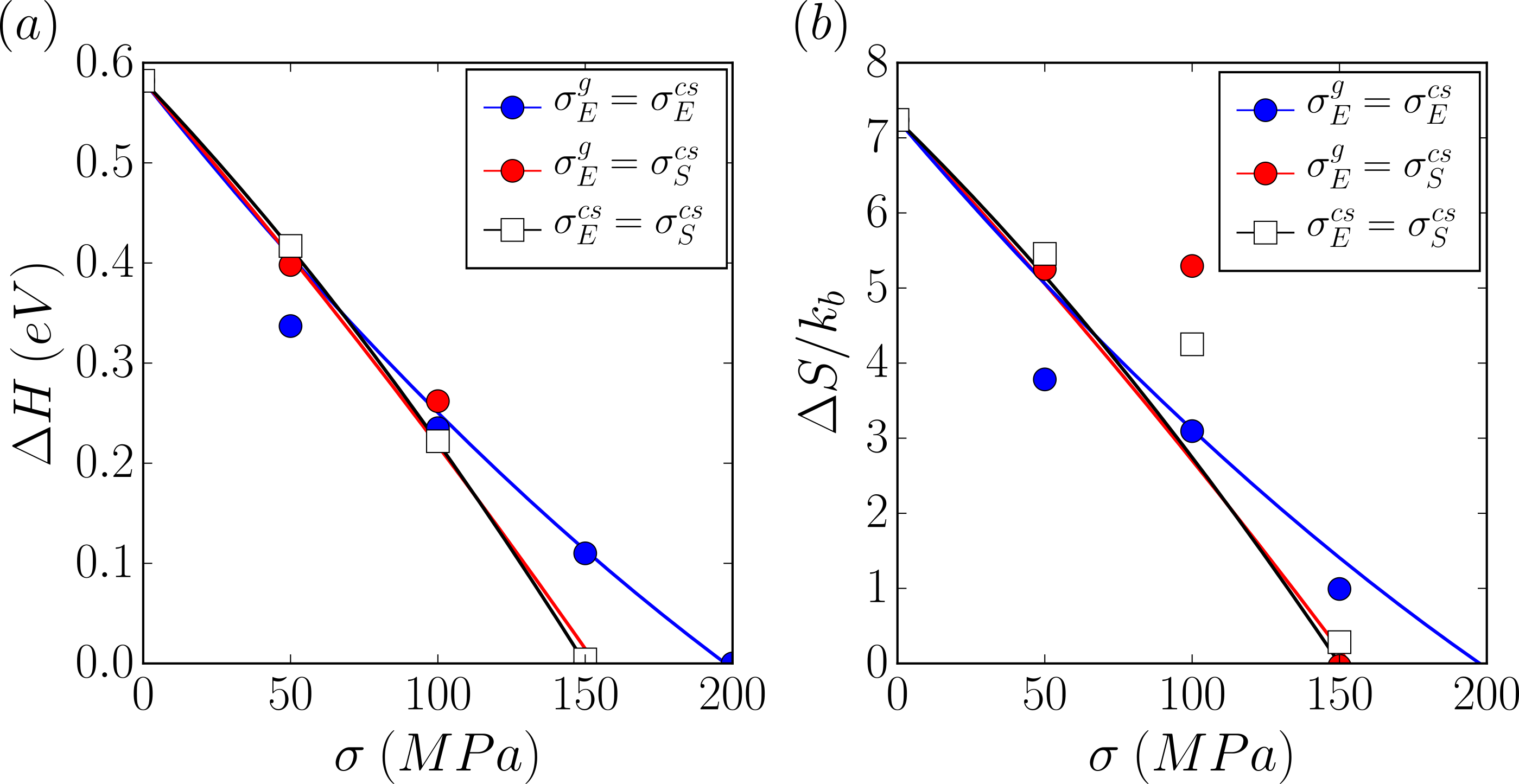}
	\caption{ $(a)$ Activation enthalpy and $(b)$ entropy barriers as a function of the applied coupled stresses ($\sigma$) for  $\sigma_E^{g}=\sigma_E^{cs}$, $\sigma_E^{g}= \sigma_S^{cs}$ and $\sigma_E^{cs}=\sigma_S^{cs}$. The lines stand for the predictions of eq. \eqref{Coup_model} and eq. \eqref{MNR}, respectively. The symbols represent the activation energies obtained from MD simulations for each pair of coupled stresses.}
	\label{Biaxial}
\end{figure}

The same strategy employed in the previous section was used to examine the effect of coupled stresses ($\sigma_E^g = \sigma_E^{cs}$,  $\sigma_E^g = \sigma_S^{cs}$ and $\sigma_E^{cs}= \sigma_S^{cs}$) on  the enthalpy and entropy barriers. The average rates as a function of $\beta$ are depicted in Fig. \ref{Coupled_rate} for the case of coupled stresses. As in the uncoupled case, the application of stress and/or temperature increases the cross-slip rate and coupled stress led to a larger reduction of the activation energy when compared to the single stress cases in Fig. \ref{Single_rate}. The dependence of the logarithm of the rates was also linear with respect to $\beta$, indicating that HTST is also a valid approximation for these stress states.  In addition, the values of $\nu_{eff}$, obtained from the rates were also inversely proportional to the applied coupled stresses, as observed in the case of uncoupled stresses, indicating that $\Delta S$ depends on the stress. The evolution of $\Delta H$ and $\Delta S$ with each pair of stresses is shown in Figs. \ref{Biaxial}$(a)$ and $(b)$, respectively. 

The interaction between coupled stress states on the free energy barrier was assumed to follow eq. \eqref{dG_expansion}. Thus, free energy barrier was reduced by the linear superposition of the individual contributions of each stress (eqs. \eqref{Uncoup_model} and \eqref{Uncoup_modelC})  which were corrected by second-order cross-terms that took into account the effects of stress coupling. Thus, 

\begin{equation}\label{Coup_model}
\begin{array}{ll}
\Delta H(\sigma_E^g+\sigma_E^{cs})= \Delta E_0 - (V_E^g\sigma_E^{g}+ V_E^{cs}\sigma_E^{cs}) - \Omega_{1}\sigma_{E}^{g}\sigma_{E}^{cs}\\
\Delta H(\sigma_E^{g}+\sigma_S^{cs})= \Delta E_0 - (V_E^g\sigma_E^{g}+ V_S^{cs}\sigma_S^{cs}) - \left[\Omega_2 \sigma_E^{cs}\sigma_S^{cs}+\Omega_{S}^{cs}(\sigma_{S}^{cs})^2\right]\\
\Delta H(\sigma_E^{cs}+\sigma_S^{cs})= \Delta E_0 - (V_E^{cs}\sigma_E^{cs}+ V_S^{cs}\sigma_S^{cs}) - \left[\Omega_3 \sigma_E^{cs}\sigma_S^{cs} + \Omega_{S}^{cs}(\sigma_{S}^{cs})^2\right]\\
\end{array}
\end{equation} 

\noindent where $V_E^{g}$, $V_E^{cs}$, $V_E^{g}$ and  the polarization factor $\Omega_S^{cs}$ were obtained from the MD simulations for uncoupled stresses (Table \ref{UncoupFit}) and $\Omega_1$, $\Omega_2$ and $\Omega_3$ (that stand for the stress interaction factors) were obtained by fitting eq. \eqref{Coup_model} to the MD results in Fig. \ref{Biaxial}. They are shown in Table \ref{CoupFit}. 

\begin{table}[H]
 \begin{center}
    \caption{Stress interaction factors (expressed in $b^3$/MPa) obtained from the MD simulations for coupled stress states.} 
    \label{CoupFit}
    \begin{tabular}{c|c|c} \hline
    \hline
       $\Omega_1$ & $\Omega_2$  & $\Omega_3$ \\
      \hline
		 -0.0258 & -0.0326 & -0.0147\\
      \hline
      \hline
    \end{tabular}
  \end{center}
\end{table}

Eqs. \eqref{Coup_model}  were able to reproduce accurately the reduction in the free energy barrier under single or coupled stress states (Fig. \ref{Biaxial}). It can be observed that they reproduce the coupled stress behavior in the case that pair stresses have the same magnitude. In addition, these models recover their original shapes when only a single stress state is considered. 
Furthermore, although the energy barrier decreases under the application of coupled stresses (as compared with the uncoupled barriers), the non-zero value of $\Omega_1$, $\Omega_2$ and $\Omega_3$ indicates first order approximation is not accurate enough and the negative sign shows that there is some screening between the Escaig and Schmid stresses. Finally, it should be noted that polarization factor $\Omega_S^{cs}$ is positive and almost twice higher in absolute value than the stress interaction factors $\Omega_1$, $\Omega_2$ and $\Omega_3$, indicating that the influence of the Schmid stress on the cross-slip plane is dominant with respect to any other stress combination at high stresses.  Furthermore, it should be noticed that the estimation of the entropic barrier combining eqs. \eqref{MNR} and \eqref{Coup_model}, with $T_m$ = 933 K, follows the trend of the MD simulation results, indicating that the MN compensation rule also holds for the biaxial case.

\subsection{Cross-slip energy barrier as a function of the stress state}

Eqs. \eqref{Coup_model}  were able to reproduce accurately the reduction in the enthalpy barrier under single or coupled stress states (Fig. \ref{Biaxial}). They can be extended using the same assumptions to a general stress state characterized by $\sigma_E^{g}$, $\sigma_E^{cs}$ and $\sigma_S^{cs}$, leading to a general expression for the energetic barrier

\begin{multline}
\Delta H(\sigma_E^g,\sigma_E^{cs},\sigma_S^{cs})= \Delta E_0 - \left(V_E^g\sigma_E^{g}+V_E^{cs}\sigma_E^{cs}+ V_S^{cs}\sigma_S^{cs}\right)- \\
 \left[ \Omega_{1}\sigma_{E}^{g}\sigma_{E}^{cs} + \Omega_{2}\sigma_{E}^{g}\sigma_{S}^{cs}  + \Omega_{3}\sigma_{E}^{cs}\sigma_{S}^{cs}  + \Omega_{S}^{cs}\left(\sigma_{S}^{cs}\right)^2 \right]
\label{Pred}
\end{multline}

\begin{figure}
	\centering
		\includegraphics[width=\textwidth]{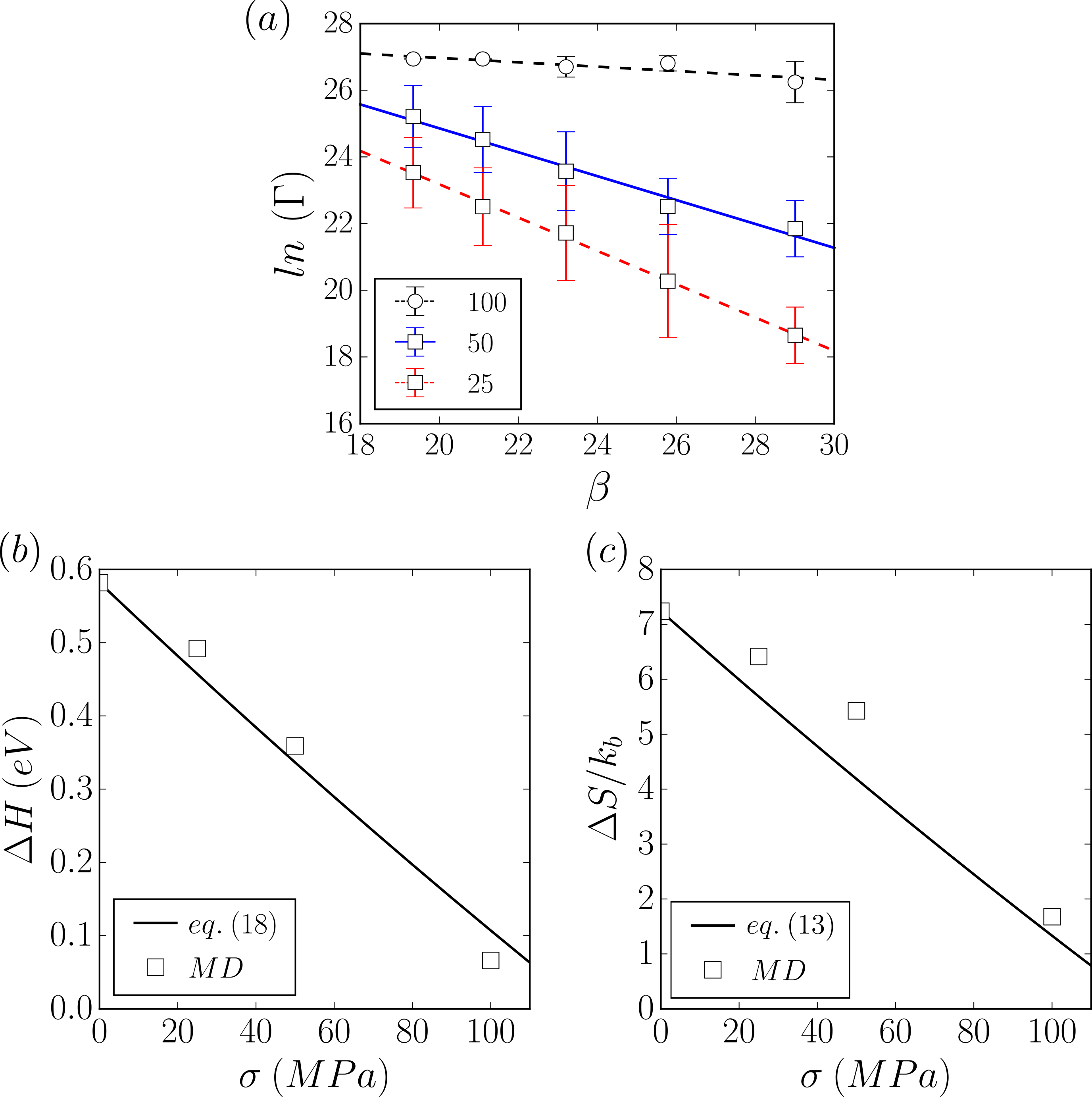}
	\caption{$(a)$ Natural logarithm of the average cross-slip rates (expressed in s$^{-1}$) as a function of $\beta$ for $\sigma = \sigma_E^{g} = \sigma_E^{cs} = \sigma_S^{cs}$.  The standard deviation from the average rate is represented by the vertical bars. $(b)$ Enthalpy $(c)$ and entropy barriers for cross-slip obtained from MD simulations (white squares) and predictions from eq. \eqref{Pred} for $\Delta H$ and from eq. \eqref{MNR} for $\Delta S$.}
	\label{Model_G}
\end{figure}

In order to evaluate the accuracy of eq. \eqref{Pred}, additional MD simulations were carried out to determine the rate of cross-slip under a multiaxial stress states with $\sigma_E^{g} = \sigma_E^{cs}= \sigma_S^{cs}$. The natural logarithm of the average rates is plotted as a function of $\beta$ in Fig. \ref{Model_G}$(a)$.  The enthalpy and entropy barriers were obtained for each stress state from the MD simualtions from the slope of straight lines and the pre-exponential factor  in this figure. They are plotted in Figs. \ref{Model_G}$(b)$ and $(c)$ as a function of the applied stress, together with the predictions  from eq. \eqref{Pred} for $\Delta H$ and from eq. \eqref{MNR} for $\Delta S$ (black lines). The agreement between the predictions of the analytical expressions and the MD simulations results is evident, showing that the model proposed is able to provide an accurate estimation of the influence of $\sigma_E^{g}$, $\sigma_E^{cs}$ and $\sigma_S^{cs}$ on the free energy barrier for cross-slip in Al. 

Probabilistic models of cross-slip have been extensively used to account for the stochastic nature of this process at the mesoscale \citep{Kubin1992, Zbib1998, Weygand2002}. They assume that the activation energy is given by an energy barrier, which depends on the applied shear stress projected in the cross-slip plane \citep{Kubin1992, Zbib1998, Weygand2002}. Hussein \textit{et al.} \citep{Hussein2015} assumed that cross-slip was only controlled by the Escaig stress along the glide and cross-slip planes and neglected the influence of the Schmid stress on the cross-slip plane. This hypothesis is not in agreement with the conclusions of our atomistic simulations  (Fig. \ref{Single_G}) and line tension models \citep{Kang2014} that demonstrate the important role of the Schmid stress on the cross-slip plane on the cross-slip rate. Martinez \textit{et al.} \citep{Martinez2008} assumed that the contribution of the applied stresses to the cross-slip rate could be calculated by a linear superposition of the Schmid stress on the cross-slip plane and of the Escaig stresses on both glide and cross-slip planes but it did not take into account the second order interactions in the case of combined Escaig and Schmid stresses along different planes, that are very important to determine the energy barrier (Fig. \ref{Biaxial}). 

Our results provide a physically-based estimation of the effect of the Schmid and Escaig stresses on the activation free energy for cross-slip rate in Al due to the polarization of the energy landscape and on  the dependence of the entropic barrier with the applied stress. Thus, a general expression of the rate of cross-slip as a function of the applied stress and temperature is obtained from MD simulations within HTST framework using MN compensation rule according to

\begin{equation}
\Gamma_{HTST} = \nu \frac{L}{L_n} e ^{-\beta \left[\Delta H(\sigma)\left(1-\frac{T}{T_m}\right)\right]}
\end{equation}
\noindent  and this methodology can be readily extended to other FCC metals.

\section{Conclusions} \label{Conclusion}
The influence of the stress state on the cross-slip rate in Al was analyzed by means of molecular dynamics simulations and transition state theory. The energy barrier in the absence of thermal energy was determined through the nudged elastic band method while the cross-slip rates were determined by means of molecular dynamics simulations for different magnitudes of the Schmid stress on the cross-slip plane, $\sigma_S^{cs}$,  and of the Escaig stresses on the cross-slip and glide planes, $\sigma_E^{cs}$ and $\sigma_E^{g}$ in the temperature range 400K-600K. Cross-slip followed the Friedel-Escaig mechanism in the whole range of stresses and temperatures explored. 

It was found that the influence of the stresses on the activation energy barrier followed the postulates of harmonic transition state theory and the enthalpy barrier and the activation entropy were determined from the average rates of cross-slip obtained from the molecular dynamics simulations. The Schmid stress on the cross-slip plane led to the largest reduction in the energy barrier at high stresses while at low stresses all stress projections showed very similar behavior. In addition, the coupling effect of the Schmid stress in the cross-slip plane and of the Escaig stresses in cross-slip and glide planes was evaluated and the synergistic contribution of all the stresses to reduce the energy barrier was determined. Based on these results, an analytical expression of the activation enthalpy for cross-slip in Al as a function of the Schmid and Escaig stresses was developed and validated for stress states involving all different combinations of the Schmid stress in the cross-slip plane and of the Escaig stresses in cross-slip and glide planes. The different terms in the analytical expressions were identified with the different contributions to the enthalpy energy barrier (configurational and due to the interaction of the applied stress with the local stress field created by the defect). Finally, it was found that the Meyer-Neldel rule is able to estimate the entropic contribution to the activation energy as a function of the applied stress. These expressions can be easily used in dislocation dynamics simulations to evaluate the probability of cross-slip of screw dislocation segments.

\section*{Acknowledgements}
This investigation was supported by the European Research Council under the European Union's Horizon 2020 research and innovation program (Advanced Grant VIRMETAL, grant agreement No. 669141). The computer resources and the technical assistance provided by the Centro de Supercomputaci\'on y Visualizaci\'on de Madrid (CeSViMa) are gratefully acknowledged. Additionally, the authors thankfully acknowledge the computer resources at Picasso and the technical support provided by Barcelona Supercomputing Center (project QCM-2018-3-0030). Finally, use of the computational resources of the Center for Nanoscale Materials, an Office of Science user facility, supported by the U.S. Department of Energy, Office of Science, Office of Basic Energy Sciences, under Contract No. DE-AC02-06CH11357, is gratefully acknowledged.


\appendix 

\setcounter{table}{0}
\setcounter{figure}{0}
\setcounter{section}{0}
\renewcommand{\thetable}{A\arabic{table}}
\renewcommand{\thefigure}{A\arabic{figure}}
\renewcommand{\thesection}{Appendix \arabic{section}}

\section{ Activation enthalpy and pre-exponential factors}\label{A1}
The activation enthalpy and the attempt frequency under different stress states and temperatures were calculated from the average rates obtained by MD according to eq. \eqref{TST_eq}. The values obtained are tabulated in the following tables:
 \begin{table}[H]
 \begin{center}
    \caption{Activation enthalpy and pre-exponential factor for uncoupled stresses.} \vskip 2truemm
    \label{}
    \begin{tabular}{c|c|c||c|c|c} \hline
    \hline 
       $\sigma_{E}^{g}$ (MPa) & $\Delta H$ (eV)  & $\nu_{eff}$ (s$^{-1}$) & $\sigma_{E}^{cs}$ (MPa) & $\Delta H$ (eV)  & $\nu_{eff}$ (s$^{-1}$) \\
      \hline
		 100 & 0.394 & 1.81$\times$10$^{13}$ & 100 & 0.429 & 2.99$\times$10$^{13}$ \\
		 150 & 0.276 & 3.39$\times$10$^{12}$ & 150 & 0.321 & 5.57$\times$10$^{12}$\\
		 200 & 0.177 & 6.44$\times$10$^{11}$ & 200 & 0.240 & 1.93$\times$10$^{12}$\\	
		 250 & 0.088 & 1.26$\times$10$^{11}$ & 250 & 0.156 & 4.39$\times$10$^{11}$\\	
      \hline
      \hline
    \end{tabular}
    \vskip 2truemm
     \begin{tabular}{c|c|c} \hline
    \hline
       $\sigma_{S}^{cs}$ & $\Delta H$ (eV)  & $\nu_{eff}$ (s$^{-1}$)\\
      \hline
		 50 & 0.496 & 4.10$\times$10$^{13}$ \\
		 100 & 0.379 & 3.06$\times$10$^{13}$\\
		 150 & 0.191 & 4.12$\times$10$^{12}$\\ 
		 200 & 0.012 & 1.70$\times$10$^{11}$\\ 
      \hline
      \hline
    \end{tabular}
  \end{center}
\end{table}

 \begin{table}[H]
 \begin{center}
    \caption{Activation enthalpy and pre-exponential factor for two stresses coupled.} \vskip 2truemm
    \begin{tabular}{c|c|c||c|c|c} \hline
    \hline
       $\sigma_{E}^{g}=\sigma_{E}^{cs}$ (MPa) & $\Delta H$ (eV)  & $\nu_{eff}$ (s$^{-1}$) & $\sigma_{E}^{g} =\sigma_{S}^{cs}$ (MPa) & $\Delta H$ (eV)  & $\nu_{eff}$ (s$^{-1}$) \\
      \hline
		 50 & 0.337 & 4.39$\times$10$^{12}$ & 50 & 0.398 & 1.91$\times$10$^{13}$ \\
		 100 & 0.235 & 2.21$\times$10$^{12}$ & 100 & 0.262 & 1.99$\times$10$^{13}$\\
		 150 & 0.110 & 2.70$\times$10$^{11}$ & 150 & 0.010 & 9.68$\times$10$^{11}$ \\	
      \hline
      \hline
    \end{tabular}
  \vskip 2truemm
    \begin{tabular}{c|c|c} \hline
    \hline
       $\sigma_{E}^{cs} =\sigma_{S}^{cs}$ (MPa) & $\Delta H$ (eV)  & $\nu_{eff}$ (s$^{-1}$)\\
      \hline
		 50 & 0.417 & 2.34$\times$10$^{13}$ \\
		 100 & 0.222 & 7.02$\times$10$^{12}$\\
		 150 & 0.004 & 1.33$\times$10$^{11}$\\ 
      \hline
      \hline
    \end{tabular}
  \end{center}
\end{table}

\begin{table}[H]
 \begin{center}
    \caption{Activation enthalpy and pre-exponential factor for the three stresses coupled.} \vskip 2truemm
    \begin{tabular}{c|c|c} \hline
    \hline
       $\sigma_{E}^{g}=\sigma_{E}^{cs}=\sigma_{S}^{cs}$  (MPa) & $\Delta H$ (eV)  & $\nu_{eff}$ (s$^{-1}$)\\
      \hline
		 25 & 0.492 & 6.10$\times$10$^{13}$\\
		 50 & 0.359 & 2.27$\times$10$^{13}$\\
		 100 & 0.066 & 5.35$\times$10$^{11}$\\	 
      \hline
      \hline
    \end{tabular}
  \end{center}
\end{table}

\section{Escaig and Schmid stress tensors applied to the atomistic domain}\label{A2}

The influence of the Escaig stresses along the glide and cross-slip plane and of the Schmid stress on the glide plane on the cross-slip rate has been evaluated by means of MD simulations. To this end, different stress states have been applied to the atomistic domain, within the NPT ensemble to attain different magnitudes of the Escaig and Schmid stresses, eq. \eqref{Esc_Sch}. They are detailed below for the different stress states (uncoupled, two stresses coupled, and three stresses coupled).  It should be noticed that the applied stress tensor is given in the coordinates  of the simulations domain ($x = [1 1 \bar{2}]$, $y = [1 1 1]$ and $z = [1 \bar{1} 0]$). 

$(a)$ \textbf{Uncoupled stresses} \\ \\
\noindent $(i)$ Escaig stress in the glide plane, $\sigma_E^g$\\
 \begin{equation}
 \sigma_{ij}  = \begin{bmatrix}
   \frac{7}{4\sqrt{2}}\sigma_E^g & -\sigma_E^g & 0 \\
   -\sigma_E^g & -\frac{7}{4\sqrt{2}}\sigma_E^g & 0 \\
   0 & 0 & 0
   \end{bmatrix}   
 \end{equation} 
 
\noindent $(ii)$ Escaig stress in the cross-slip plane, $\sigma_E^{cs}$\\
 \begin{equation}
  \sigma_{ij}  = \begin{bmatrix}
   \frac{9}{4\sqrt{2}}\sigma_E^{cs} & 0 & 0 \\
   0 & -\frac{9}{4\sqrt{2}}\sigma_E^{cs} & 0 \\
   0 & 0 & 0
   \end{bmatrix}
 \end{equation} 
 
\noindent $(iii)$ Schmid stress in the cross-slip plane, $\sigma_S^{cs}$\\
  \begin{equation}
  \sigma_{ij} = \begin{bmatrix}
   0 & 0 & 3\sigma_{S}^{cs} \\
   0 & 0 & 0 \\
   3\sigma_{S}^{cs} & 0 & 0
   \end{bmatrix}
   \end{equation}\\ \\
   
 $(b)$ \textbf{Two stresses coupled}  \\ \\
 \noindent $(i)$ Escaig stresses  in the cross-slip  and glide planes, $\sigma_E^{cs}$ and $\sigma_E^{g}$ \\
    \begin{equation}
   \sigma_{ij} = \begin{bmatrix}
   \frac{9}{4\sqrt{2}}\sigma_E^g+\frac{7}{4\sqrt{2}}\sigma_E^{cs}  & -\sigma_E^g & 0 \\
   -\sigma_E^g & -\frac{9}{4\sqrt{2}}\sigma_E^g-\frac{7}{4\sqrt{2}}\sigma_E^{cs} & 0 \\
   0 & 0 & 0
   \end{bmatrix}
   \end{equation} 
   
 \noindent $(ii)$ Escaig and Schmid stresses in the cross-slip plane, $\sigma_E^{cs}$ and $\sigma_S^{cs}$
    \begin{equation} 
 \sigma_{ij}  = \begin{bmatrix}
   \frac{9}{4\sqrt{2}}\sigma_E^{cs} & 0 & 3\sigma_{S}^{cs} \\
   0 & -\frac{9}{4\sqrt{2}}\sigma_E^{cs} & 0 \\
   3\sigma_{S}^{cs} & 0 & 0
   \end{bmatrix}
      \end{equation}
      
  \noindent  $(iii)$ Escaig stress in the glide plane and Schmid stress in the cross-slip plane, $\sigma_E^{g}$ and $\sigma_S^{cs}$\\
   \begin{equation}
    \sigma_{ij} = \begin{bmatrix}
  \frac{7}{4\sqrt{2}}\sigma_E^g & -\sigma_E^g & 3\sigma_{S}^{cs} \\
   -\sigma_E^g & -\frac{7}{4\sqrt{2}}\sigma_E^g & 0 \\
   3\sigma_{S}^{cs} & 0 & 0
   \end{bmatrix}
   \end{equation}\\ \\
$(c)$ \textbf{Three stresses coupled}  \\

\noindent Escaig stresses  in the cross-slip  and glide planes and Schmid stress in the cross-slip plane, $\sigma_E^{cs}$, $\sigma_E^{g}$ and $\sigma_S^{cs}$ \\
   \begin{equation}
   \sigma_{ij} = \begin{bmatrix}
   \frac{9}{4\sqrt{2}}\sigma_E^g+\frac{7}{4\sqrt{2}}\sigma_E^{cs}  & -\sigma_E^g & 3\sigma_{S}^{cs} \\
   -\sigma_E^g & -\frac{9}{4\sqrt{2}}\sigma_E^g-\frac{7}{4\sqrt{2}}\sigma_E^{cs}  & 0 \\
   3\sigma_{S}^{cs} & 0 & 0 
      \end{bmatrix}
	\end{equation}

\end{document}